# Effect of Al content on the critical resolved shear stress for twin nucleation and growth in Mg alloys

Jingya Wang[a, b], Jon M. Molina-Aldareguía[a], Javier LLorca[a, b,*]

[a]IMDEA Materials Institute, C/Eric Kandel 2, 28906, Getafe, Madrid, Spain
[b]Department of Materials Science, Polytechnic University of Madrid/Universidad Politécnica de Madrid, E.T.S. de Ingenieros de Caminos, 28040 Madrid, Spain

**Abstract**

The effect of Al atoms in solid solution on the critical resolved shear stress for twin nucleation and growth was analyzed by means of the combination of diffusion couples with compression tests in micropillars oriented for twinning. The critical resolved shear stress for twin nucleation was higher than that for twin growth and both increased by the same amount with the Al content. Nevertheless, the increase was small (≈ 10 MPa) for 4 at.%Al but large (up to 60-70 MPa) for 9 at.%Al. These results were in agreement with Labusch-models based on first principles calculations in the dilute regime (< 5 at.%Al) [51]. Comparison with recent data in the literature showed that Al atoms are more effective in increasing the critical resolved shear stresses for twin nucleation and growth than for basal slip [21]. Finally, compression tests in micropillars oriented along [0001] showed the critical shear stress for pyramidal slip increased rapidly with the Al content from 98 MPa in pure Mg to 250 MPa in Mg-9 at.%Al. Thus, the addition of Al increased the plastic anisotropy of Mg alloys.

* Corresponding author: javier.llorca@imdea.org (J. LLorca)

# 1. Introduction

Magnesium and its alloys have attracted significant interest recently for engineering applications in automotive, aerospace, electronics, and biomedical sectors due to their low density, high specific-stiffness, superior damping capacity and biocompatibility [1–5]. Mg alloys have a hexagonal closed packed (HCP) lattice structure and plastic deformation in Mg alloys tends to be localized along the basal plane because the critical resolved shear stress (CRSS) for <$a$> basal slip is around ~0.5 MPa [6–8]. Deformation perpendicular to the basal plane cannot be easily accommodated by <$c+a$> pyramidal slip at room temperature because the CRSS for pyramidal slip is very high (~80 MPa [9]) and tensile twinning $\{10\bar{1}2\}<10\bar{1}1>$, that is activated at low values of the CRSS (~12 MPa [9]), becomes the second dominant plastic deformation mechanism. Pure Mg is very soft as a result of the low CRSS values for basal slip and twinning and, in addition, twinning is a polar deformation mechanism that is only activated when leading to an extension of the $c$ axis of the HCP lattice. As a result, textured Mg alloys show a strong plastic anisotropy which leads to poor ductility and formability at room temperature, hindering their application in structural elements. Thus, understanding and controlling twinning is of crucial importance to design Mg alloys with improved properties.

Twinning deformation mainly comprises two different processes: twin nucleation and twin growth (thickening). Twin nucleation is heterogeneous and takes place in regions with large stress concentrations in the microstructure, such as grain boundaries [10,11]. Several theories were proposed in the past for twin nucleation, including the pole mechanism of Thompson and Millard [12] and the slip dislocation dissociation mechanism of Mendelson [13]. More recently, twin nucleation has been studied by means of atomistic simulations [10,14–16] and *in situ* mechanical tests on micropillars [17] and these studies have revealed several dislocation-based structures of twin nuclei. Regarding the twin growth, it is generally accepted that twin thickening is mediated by the glide of twinning dislocations along the twin planes [18], and such migration is controlled by the resolved shear stress on the twin plane and direction.

Tailoring the mechanical properties of Mg alloys by means of alloying is an obvious strategy that requires a detailed knowledge of the effect of solute atoms on the CRSS of the different plastic deformation mechanisms. This has been achieved for basal slip in Mg-Al and Mg-Zn alloys by means of tests in single crystals [19,20] and micropillar compression tests [21]. Regarding solid solution strengthening of tensile twinning, it has been reported that Y, Zr, Nd and Li increase the CRSS for twinning [22–25], while the effect of Zn is negligible [26,27]. For instance, the CRSS value to activate tensile twinning in Mg was increased up to ~100 MPa by the addition of 10 at.%Y [23], while 5 wt.%Li improved the CRSS for twinning up to ~ 25 MPa [25]. However, these analyses were carried out in polycrystalline samples and the effect of the alloying elements on the CRSS for twinning was estimated indirectly after it was isolated from other hardening mechanisms (grain boundaries, latent hardening and the influence of the solute atoms on basal, prismatic and pyramidal slip). Moreover, these strategies based on tests in polycrystals cannot distinguish between the effect of solute atoms on the CRSS for twin nucleation and growth.

These limitations can be overcome through the application of micromechanical testing techniques. In fact, micropillar compression tests have been extensively used to analyse the different deformation mechanisms of Mg and Mg alloys along specific orientations [17,21,26,28–35], overcoming the limitations of tests in polycrystals. For instance, micropillar compression tests were performed in pure Mg along the $a$-axis [17,26,28,30,31,35] to explore the effect of the micropillar size on the mechanisms of $\{10\bar{1}2\}$ twin deformation and on the CRSS for twin growth. Using these techniques, they estimated CRSSs for twin growth of ~7 MPa [26] and 29 MPa [35] in pure Mg.

In this investigation, micropillar compression tests are used in combination with diffusion couples to analyze the mechanisms of twin nucleation and growth in Mg-Al alloys. To this end, micropillars of cross-sections in the range 3x3 $\mu m^2$ to 7x7 $\mu m^2$ were milled from large grains with different Al content in the diffusion couples and were tested in compression along the $[01\bar{1}0]$ orientation. The CRSS values for twin nucleation and growth could be obtained from the stress-strain curves after careful examination of the deformation mechanisms and of the effect of micropillar size. This information, together with the CRSS for basal slip from a previous investigation [21] and for pyramidal slip obtained from micropillar compression tests along [0001], showed that the plastic anisotropy of Mg-Al alloys increased with the Al content.

## 2. Material and experimental procedures

Pure Mg and Mg-9 at.%Al alloys were melted and cast in an induction furnace in Ar from high-purity Mg (99.99 wt.%) and Al (99.99 wt.%) pellets. The cast ingots were homogenized in Ar within quartz capsules at 673 K for 15 days and discs of 12 mm in diameter and 7.5 mm in length were machined. The surfaces of the pure Mg and Mg-9 at.%Al discs were polished and diffusion-bonded in vacuum under a compression force of 800 N for 1 h at 673 K to create a Mg/Mg-9 at.%Al diffusion couple, which was annealed in Ar at 673 K during 352 h to obtain the composition gradient and release the thermal stresses. More details of the process to manufacture the diffusion couple can be found in [21].

Square samples of 7x7 $mm^2$ and 2 mm in thickness were cut from the diffusion couple perpendicular to the bonding interface and chemically polished to remove any residual surface damage. Afterwards, the composition profiles perpendicular to the bonding interface within the diffusion region were measured by the electron probe microanalysis (EPMA) using Wavelength Dispersive Spectroscopy (WDS) with a voltage of 20 kV and a beam current of 50 nA in a JEOL Superprobe JXA-8900M. In addition, the microstructure of the specimens was characterized using electron backscatter diffraction (EBSD) in a dual beam scanning electron microscope (SEM) Helios Nanolab 600i FEI at an accelerating voltage of 30 kV and a beam current of 2.7 nA to identify the grain orientation and the grain size. The grain sizes within the diffusion region were very large (> 1 mm), as shown in [21].

Micropillars with a square cross-section with selected chemical compositions (pure Mg, Mg-4 at.%Al and Mg-9 at.%Al) were fabricated within the interdiffusion region in the centre of grains oriented perpendicularly to the $[01\bar{1}0]$ orientation. The cross section of the micropillars was varied from 3x3 $\mu m^2$ to 5x5 $\mu m^2$ to 7x7 $\mu m^2$ to explore the effect of the size on the mechanical behaviour. The aspect ratio of the micropillars was in the range 2:1 to 3:1 to avoid buckling while the stress state was uniform along most of the length of the micropillar (Figure 1a) [36,37]. The micropillars were fabricated using a FEI Helios Nanolab 600i FEI focused ion beam (FIB) microscope with a $Ga^+$ ion beam operated at 30 kV. The initial beam current was 9.3 nA and it was reduced to 40 pA in the final polishing step to minimize the surface damage due to the $Ga^+$ ion implantation. Moreover, the sample was tilted ±2° with respect to the ion beam axis during the final milling to reduce the taper, which was always below 1º (Figure 1a).

Micropillar compression tests were carried out using a TI950 Triboindenter (Hysitron, Inc., Minneapolis, MN) equipped with a diamond flat punch with a diameter of 15 μm. All the tests were carried out at a constant strain rate of $10^{-3}$ $s^{-1}$ under the displacement control up to a maximum strain of 10%. The load-displacement curves were corrected by applying the Sneddon method [38,39] to account for the compliance associated with the elastic deflection of the base of the micropillar. The engineering stress-strain curves were obtained from the corrected curves from the cross-sectional area and the height of the micropillar before deformation. At least three tests were repeated for each condition (alloy composition and micropillar size) to ensure the reproducibility of the results. It has been shown that the mechanical properties measured with square-section micropillars are equivalent to those obtained with circular ones [40]. It should be noted that ex situ micropillar compression tests were selected to analyse the mechanical behaviour instead of in situ tests [21] because the only additional information provided by the in situ tests (with respect to the ex situ tests) is the continuous imaging of the two external surfaces of the micropillar. But twin nucleation and growth may occur within the micropillar without modifying the observed external surfaces.

After the compression tests, the morphology of the deformed micropillars was carefully examined using secondary electrons in the SEM. In addition, transmission Kikuchi diffractiob (TKD) and transmission electron microscopy (TEM) were used to characterize the deformation mechanisms for selected micropillars. To this end, thin lamella of the cross-sections of the deformed micropillars were extracted using the standard lift-out technique [41] in the FEI Helios 600 FIB microscope at 30 kV. The lamella was thinned to approximately 100 nm by adopting a range of ion beam currents from 2.5 nA to 40 pA and finally polished at 5 kV with an ion beam current of 15 pA to minimize the $Ga^+$ ion beam damage. The TKD mapping was carried out in the Helios Nanolab 600i FEI microscope equipped with an Oxford Instruments NordlysNano EBSD detector. The electron beam was operated at 15 kV with a beam current of 2.7 nA and the TKD maps were collected with a step size of 50 nm. The TEM observations were performed in a Talos F200X microscope at an accelerating voltage of 200 kV.

## 3. Results

### 3.1 Deformation mechanisms of pure Mg during [01$\bar{1}$0] compression

The micropillars were deformed parallel to the [01$\bar{1}$0] orientation of the Mg lattice. The experimental values of the CRSS of the different deformation modes of pure Mg under compression in the literature [6–9,42,43] are summarized in Table 1, together with the corresponding Schmid factors (*m*) for compression along the [01$\bar{1}$0], [01$\bar{1}$6] and [0001] orientations. The second orientation corresponds to the lattice orientation after $\{10\bar{1}2\}$ tensile twinning, while [0001] direction was selected to study pyramidal slip. The Schmid factor for all the slip systems and tensile twinning were computed based on the crystal orientation information obtained from the EBSD measurements following the protocol in [44].

Although the CRSS for basal slip is very low (≈0.5 MPa), this mechanism is not expected to be dominant in the micropillar oriented along the $[01\bar{1}0]$ orientation because the Schmid factor for basal slip is almost zero. $\{10\bar{1}2\}$ tensile twinning shows the highest Schmid factor (0.48 - 0.49), followed by prismatic slip (0.46). However, the CRSS for prismatic slip is much higher than that of tensile twinning and, therefore, $\{10\bar{1}2\}$ tensile twinning is expected to be the preferred deformation mode under $[01\bar{1}0]$ compression. It is worth noting that two twin variants have very high Schmid factors and, thus, they have similar probabilities to develop during deformation.

After twinning, the loading axis changes to the $[01\bar{1}6]$ orientation in the twinned region as a result of the crystal reorientation by ~86°, which is schematically shown in Fig. 1b. The new compression direction has a misorientation ~5° with respect to [0001] crystal orientation [26,30], and the Schmid factors for all deformation modes in the twinned region are listed in Table 1. Since the new compression axis is quite close to the [0001] orientation, the active deformation mechanisms may be expected to be similar to the ones found during compression along the [0001] orientation [30,32,33].

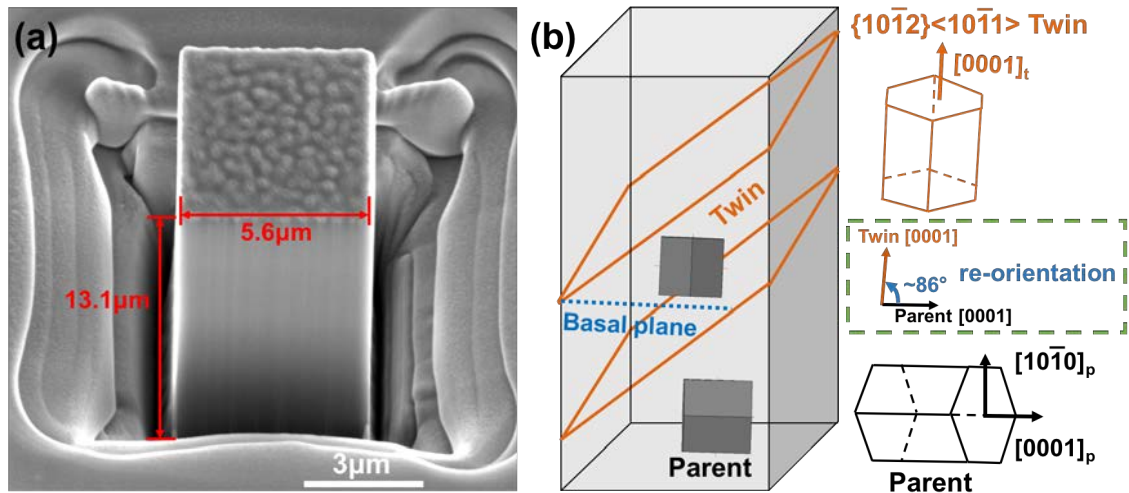

**Fig. 1.** (a) Representative micropillar of 5 x 5 μm² cross-section. (b) Schematic of the orientation relationship between the parent and twin regions in the micropillar. The parent crystal is reoriented by ~86° by twinning.

A representative engineering stress-strain curve is plotted in Fig. 2 from the compression test of a micropillar of pure Mg of 5x5 µm$^2$ cross-section. The initial linear elastic response is abruptly finished by an abrupt load drop followed by a strain burst. Afterwards, the deformation progresses at constant stress up to an applied strain of 5.5% when the micropillar shows a very high strain hardening up to the maximum applied strain of 10%. The different regions in the stress-strain curve are expected to correspond to different dominant deformation mechanisms and, thus, various tests were stopped after the first load drop ($\varepsilon \approx 1\%$), at the end of the plateau region ($\varepsilon \approx 6\%$) and at $\varepsilon \approx 10\%$ to analyze the microstructure. The corresponding points are marked with (b), (c) and (d), respectively, in Fig. 2a.

**Table 1**. Critical resolved shear stress (CRSS) and Schmid factor (*m*) of the possible deformation modes in Mg under compression along the [01$\bar{1}$0], [01$\bar{1}$6] and [0001] orientations. The [01$\bar{1}$6] orientation corresponds to the crystal orientation after twinning.

| Deformation mode | | CRSS (MPa) | Schmid factor (*m*) | | |
|---|---|---|---|---|---|
| | | | [01$\bar{1}$0] | [01$\bar{1}$6] | [0001] |
| Basal | (0001)[11$\bar{2}$0] | 0.5 [6–8] | 0 | 0.13 | 0.03 |
| | (0001)[1$\bar{2}$10] | | 0 | 0.12 | 0.05 |
| | (0001)[$\bar{2}$110] | | 0 | 0.02 | 0.02 |
| Prismatic | (1$\bar{1}$00)[11$\bar{2}$0] | 39 [42] | 0.46 | 0.01 | 0 |
| | (10$\bar{1}$0)[1$\bar{2}$10] | | 0.39 | 0.01 | 0 |
| | (01$\bar{1}$0)[$\bar{2}$110] | | 0.07 | 0 | 0 |
| Pyramidal II | (11$\bar{2}$2)[$\bar{1}\bar{1}$23] | 80 [9] | 0.27 | 0.49 | 0.43 |
| | ($\bar{1}$2$\bar{1}$2)[1$\bar{2}$13] | | 0.32 | 0.48 | 0.43 |
| | (1$\bar{2}$12)[$\bar{1}$2$\bar{1}$3] | | 0.39 | 0.38 | 0.47 |
| | ($\bar{1}\bar{1}$22)[11$\bar{2}$3] | | 0.34 | 0.37 | 0.46 |
| | (2$\bar{1}\bar{1}$2)[$\bar{2}$113] | | 0 | 0.44 | 0.45 |
| | ($\bar{2}$112)[2$\bar{1}\bar{1}$3] | | 0.37 | 0.07 | 0.16 |
| {10$\bar{1}$2} twin | (10$\bar{1}$2)[$\bar{1}$011] | 12 [43] | 0.09 | 0.48 | -0.49 |
| | (1$\bar{1}$02)[$\bar{1}$101] | | 0.15 | 0.49 | -0.49 |
| | (01$\bar{1}$2)[0$\bar{1}$11] | | 0.49 | 0.47 | -0.50 |
| | (0$\bar{1}$12)[01$\bar{1}$1] | | 0.48 | 0.49 | -0.49 |
| | ($\bar{1}$012)[10$\bar{1}$1] | | 0.09 | 0.49 | -0.49 |
| | ($\bar{1}$102)[1$\bar{1}$01] | | 0.15 | 0.48 | -0.50 |

The observation of the micropillar in the SEM after the sudden load drop at 1% strain (Fig. 2b) showed a slight tilt along the edge of the lateral surface (marked with an arrow). A thin lamella was extracted from the deformed micropillar and characterized by TKD. This analysis showed that the two variants of the {10$\bar{1}$2} twin (T1 and T2) with the highest Schmid factor were nucleated in the upper part of the micropillar (Fig. 2e), leading to a ~ 86° rotation from the parent lattice. The crystal orientation of both twinning variants and of the parent are schematically shown in Fig. 2g. The twin boundary (TB) and the boundary between the two twin variants (TTB) were marked by black and yellow lines, respectively, in Fig 2(e). Thus, the load drop was associated with the crystal reorientation induced by the nucleation of two $\{10\bar{1}2\}$ twin variants.

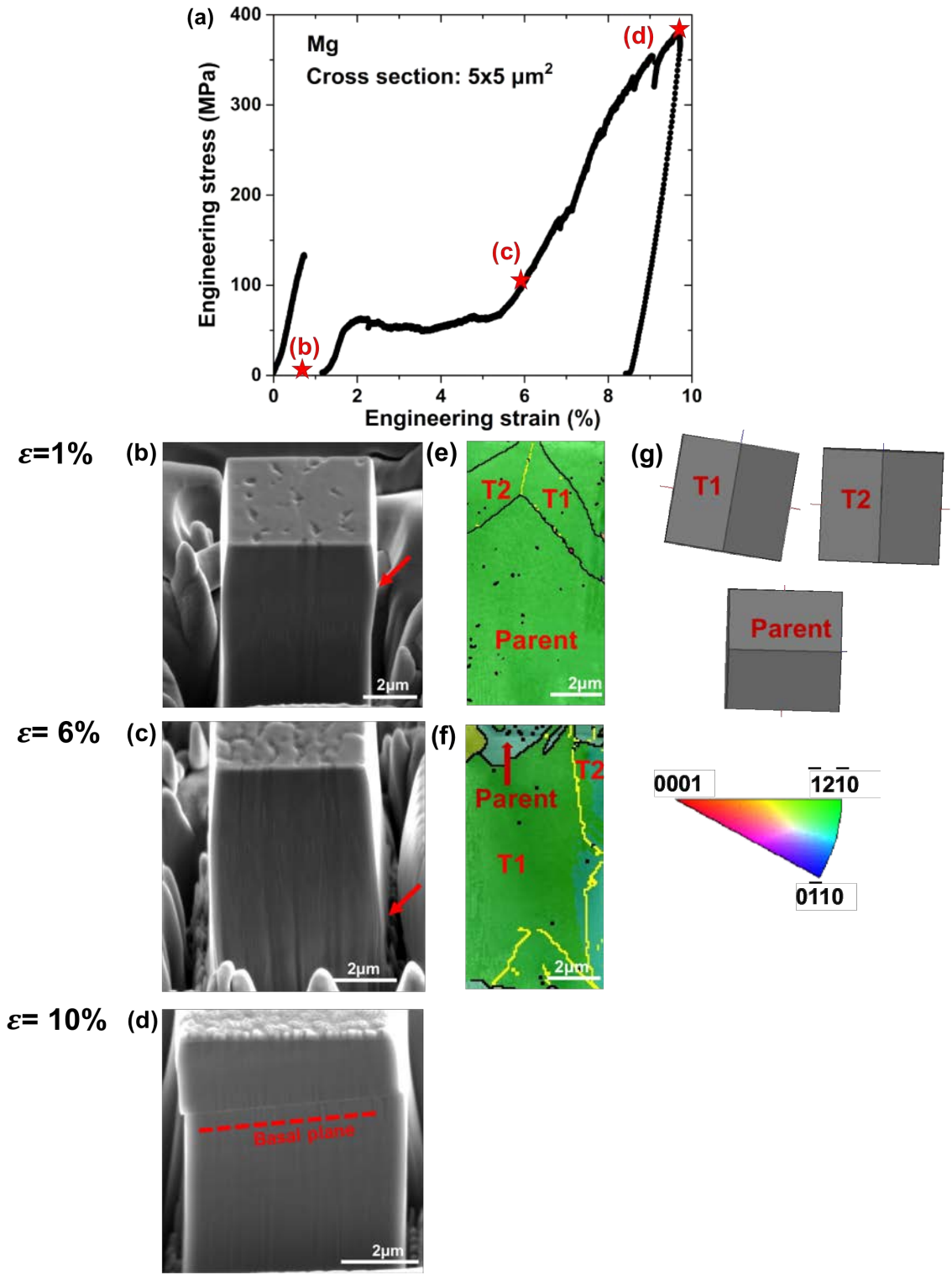

Fig. 2. (a) Representative engineering stress-strain curve of a pure Mg micropillar of 5x5 μm$^2$ deformed in compression along the [01$\bar{1}$0] orientation. (b) SEM micrograph of the deformed micropillar after the load drop at ~ 1% strain. (c) *Idem* at 6% strain. (d) *Idem* at 10% strain. (e-f) *t*-EBSD maps of a thin lamella extracted from the deformed micropillars in (b-c). (g) Schematic of the HCP lattice orientation in both twin variants and in the parent crystal. Black and yellow lines in (e-f) stand for the twin boundary (TB) and the boundary between the two twin variants (TTB).

After twin nucleation, the flow stress remained constant up to an applied strain of ≈5.5%. The SEM micrograph of another micropillar deformed up to 6% shows that the tilt feature along the lateral edge has propagated up to the bottom of the micropillar (Fig. 2c) and the orientation map obtained by TKD of another lamella extracted from the micropillar is shown in Fig. 2f. It confirmed that the whole micropillar was twinned by the thickening of both twin variants that took place at constant stress.

Another micropillar was tested up to an applied strain of 10%, and the SEM micrograph of the deformed micropillar showed the presence of slip traces parallel to the basal plane of the twinned crystal on the lateral surface (Fig. 2d). Although the orientation within the twinned region is not favourably oriented for basal slip (~ 8° misorientation with respect to the *c*-axis), slip traces were found on the lateral surface of the micropillar (Fig. 2d) because the applied stress was very large (close to 400 MPa) and the CRSS for basal slip is quite low.

In order to further investigate the deformation mechanisms, the lamella extracted from the micropillars deformed up 6% and 10% were examined by TEM. A reduced number of <*a*> dislocations are observed within the parent crystal on the prismatic plane in the micropillar deformed up to 6%, as shown in Fig. 3a under the diffraction condition of *g*=[01$\bar{1}$1]$_p$. Although the CRSS value for prismatic slip is very high (Table 1), it is possible that prismatic slip was activated taking into account the compressive stress attained to nucleate the twin and the Schmid factor of some prismatic slip systems [30,44]. Nevertheless, their contribution to the plastic slip should be small taken into account the small density of these dislocations. A few dislocations were also observed on non-basal planes in the vicinity of the twin boundary within the twinned region with a <*c*> component when the applied strain was 6% (Fig. 3b). At the *g*=[0002]$_t$ diffraction condition, the visible dislocations have a Burgers vector with a <*c*> component based on the dislocation extinction criterion, and they may be <*c*> or <*c+a*> dislocations. The density of these dislocations increased dramatically when the applied strain reached 10%, as shown in Fig. 3c. When the diffraction condition is changed to *g*=[01$\bar{1}$1]$_t$, dislocations containing an <i><a></i> component could be revealed and more dislocations are observed under $g=[01\bar{1}1]_t$ in Fig. 3d, confirming the presence of the dislocations with Burgers vector <i><c+a></i>. The large density of <i><c+a></i> dislocations in the prismatic plane within the twinned region, associated to the large strain hardening of the stress-strain curve, indicates that pyramidal slip was progressively activated within the twinned micropillar for applied strains > 6% and become the dominant plastic deformation mechanism once the whole micropillar was twinned. This statement will be supported by additional evidence presented in section 4.3.

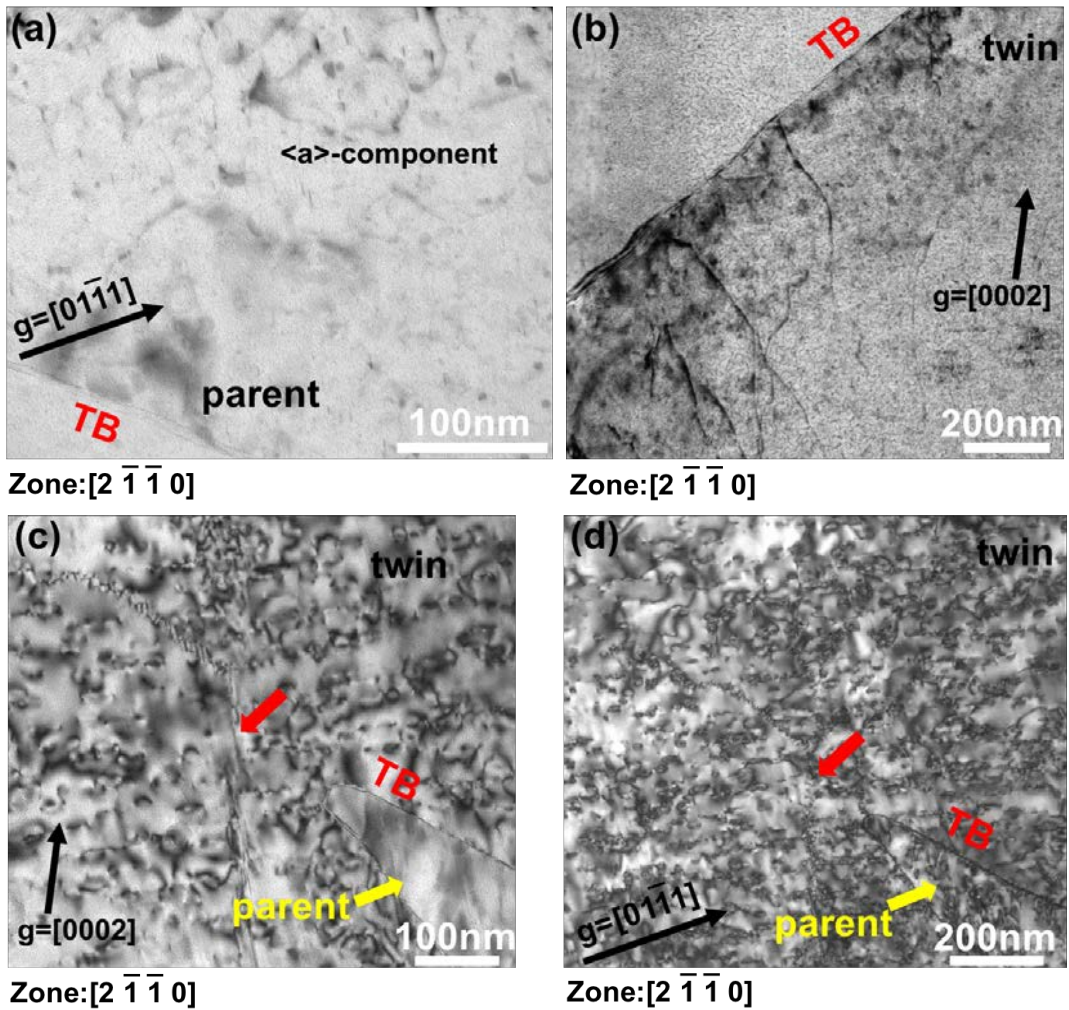

Fig. 3. TEM micrographs of lamella extracted from micropillars deformed up to different strains. (a) Applied strain 6%. <i><a></i> dislocations in the prismatic plane within the parent region, $g=[01\bar{1}1]_p$. (b) Applied strain 6%. <i><c></i> dislocations in the twinned region, $g=[0002]_t$. (c) Applied strain 10%. <i><c></i> dislocations in the twinned region, $g=[0002]_t$. (d) Same as (c) but observed under the $g=[01\bar{1}1]_t$, confirming the presence of <i><a+c></i> dislocations in this region. The zone axis was $[2\bar{1}\bar{1}0]$ in all cases.

In this investigation, the twin nucleation stress was determined from the critical point of load drop (associated with the formation of twin), while the twin growth stress was corresponding to the stress-plateau that was associated with the migration of the twin boundaries, as stated above. The magnitude of the critical stresses for twin nucleation and growth are consistent with the results of Wang. et al. [26], who tested Mg micropillars in compression oriented for twinning. Some of the micropillars were twin-free while others contained twins and, thus, the critical stresses for twin nucleation and growth were determined in different micropillars. In our case, both magnitudes were determined from a single test in a twin-free micropillar.

### 3.2 Deformation mechanisms of Mg-Al alloys during [$01\bar{1}0$] compression

A representative engineering stress-strain curve of a micropillar of 7x7 μm$^2$ is plotted in Fig. 4a for a Mg-9 at.%Al alloy. The micropillar was milled in another grain with the same [$01\bar{1}0$] oreintation and the general features of the curve were similar to those reported for pure Mg (Fig. 2a). A pronounced load drop took place at 1% elastic strain and was accompanied by a strain burst up to 3% strain which was associated with twin nucleation at the upper region of the micropillar. Deformation progressed at constant stress afterwards, followed by a step hardening region for strains > 5%. The stress carried by the micropillar at 10% was close to 600 MPa, much higher than the one in the case of pure Mg (≈ 400 MPa). Moreover, strain bursts (that can be noticed by sudden drops in the load during deformation) were found in the plateau and hardening regions of the Mg-9 at.%Al alloy but not in the case of the pure Mg micropillars.

Micropillar compression tests were stopped at 1% strain, after the sudden load drop, and at 10% strain to further analyze the deformation mechanisms. The micropillar deformed up to 1% showed a slight tilt along the edge of the lateral surface (Fig. 4b), as in the case of pure Mg. A thin lamella was extracted from the micropillar and characterized by TEM. The twinned region occupied the upper part of the whole pillar, and two TBs can be found in Fig. 4d. They were parallel to the twin plane, indicated by the red dotted line. Moreover, the TTB between the two {$10\bar{1}2$} twin variants (T1 and T2) is marked by the yellow dashed line in Fig. 4d. The presence of the twin and of the parent were explicitly confirmed by the diffraction patterns.

The lateral surface of the micropillar deformed up to 10% also showed slip traces parallel to the basal plane of the twinned crystal (Fig. 4c). The TKD image of a lamella extracted from this micropillar showed that the whole micropillar has been twinned and one of the twin variants has propagated and replaced the other variant (Fig. 4e), which was only found at the bottom-right corner of the micropillar. The crystal orientation of the twin variants and of the parent are shown in Fig. 4f. Similar to the case of pure Mg, one twin variant grew at the expense of the other variant during deformation. The slip traces and the slip steps corresponding to the basal plane in the twinned crystal were more marked than in pure Mg (Fig. 4c) and they were responsible for the strain bursts observed in Fig. 4a during strain hardening.

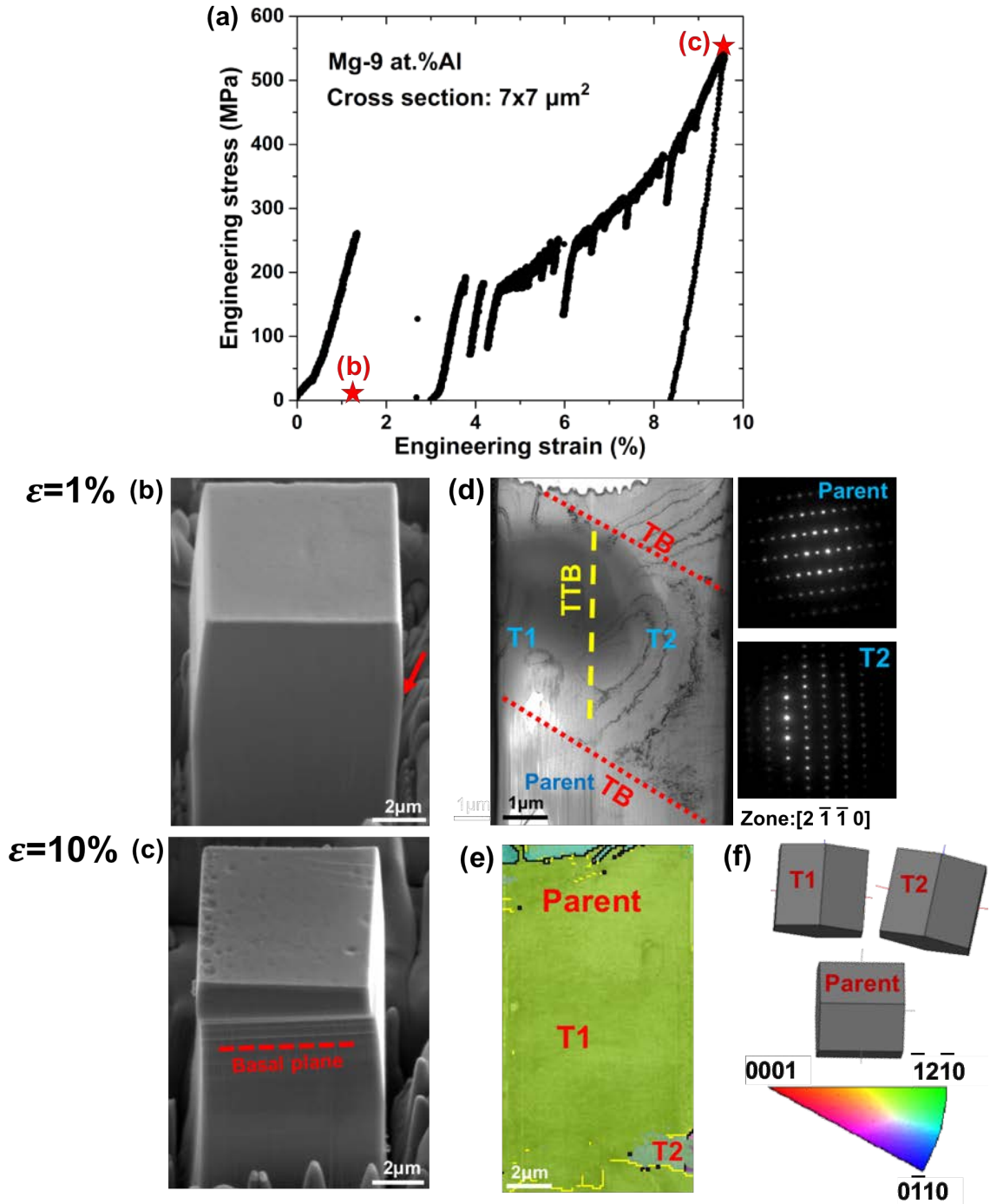

Fig. 4 (a) Representative engineering stress-strain curve of a Mg-9 at.%Al micropillar of 7x7 µm² deformed in compression along the [01$\bar{1}$0] orientation. (b) SEM micrograph of the deformed micropillar after the load drop at ~ 1% strain. (c) *Idem* at 10% strain. (d) TEM micrograph of the thin lamella extracted from the micropillar in (b), including the diffraction patterns of the parent and twin regions. (e) TKD maps of a thin lamella extracted from the micropillar in (c). Black and yellow lines stand for the twin boundary (TB) and the boundary between the two twin variants (TTB). (f) Schematic of the HCP lattice orientation in both twin variants and in the parent crystal.

The dislocation structure within the parent and twinned regions was analysed by TEM. Prismatic <a> dislocations in the parent region (Fig. 5a) as well as pyramidal <c+a> dislocations in the twinned region (Fig. 5b) were found in the micropillar deformed up to 1%, as in the case of pure Mg. The density of pyramidal <c+a> dislocations in the twinned region increased when the micropillar was deformed up to 10% (Fig. 5c) but the dislocation density was significantly lower than in the case of pure Mg. These differences are obviously related to the higher strain hardening of the Mg-9 at.%Al micropillars and they will be analysed in more detail below.

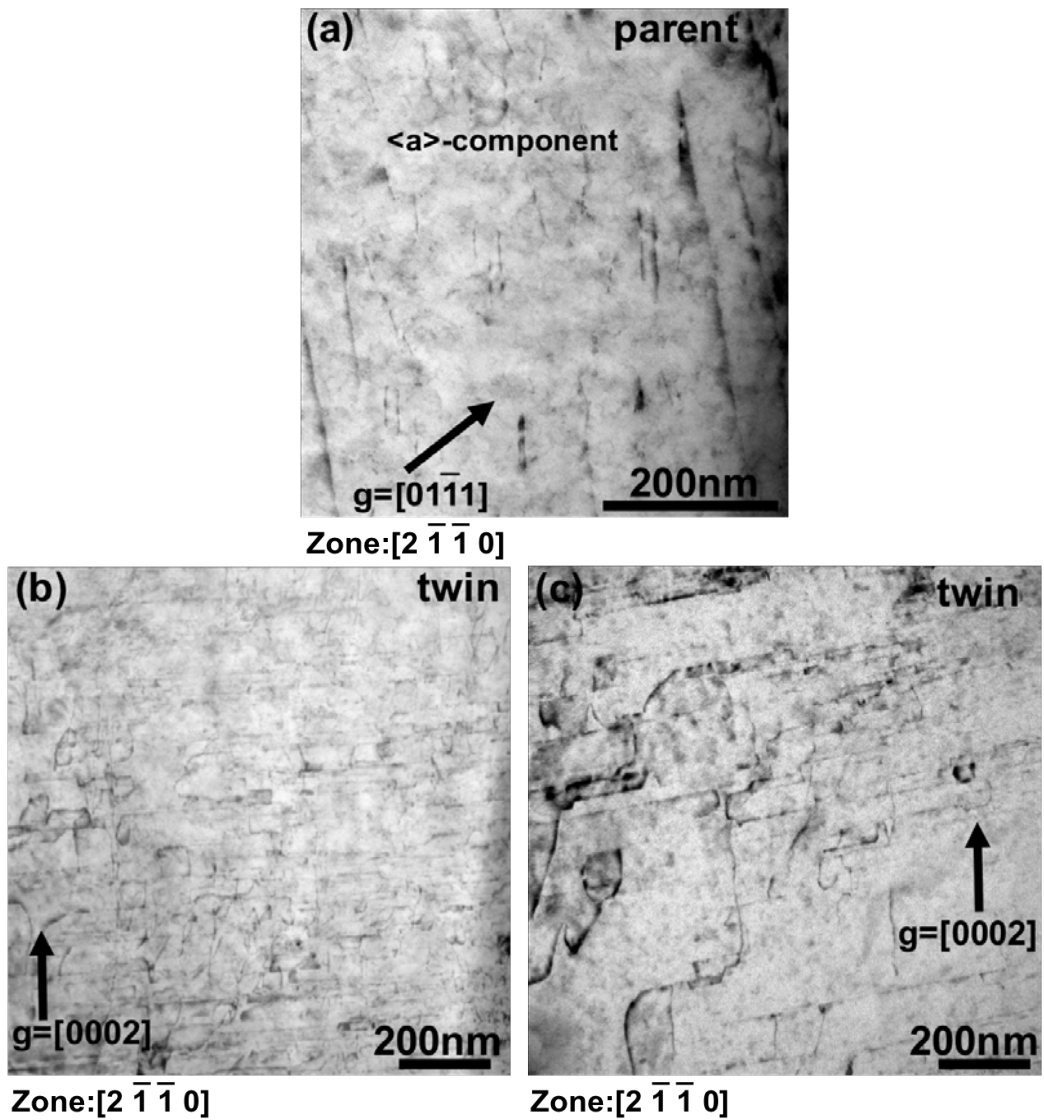

Fig. 5. (a) Applied strain 1%. <a> dislocations in the prismatic plane within the parent region, $g=[01\bar{1}1]_p$. (b) Applied strain 1%. <c> dislocations in the twinned region, $g=[0002]_t$. (c) Applied

strain 10%. <c> dislocations in the twinned region, g=[0002]$_t$. The zone axis was [2$\bar{1}$10] in all cases.

## 3.3 Size effects during [01$\bar{1}$0] compression of Mg and Mg-Al alloys

The engineering stress-strain curves of the compression tests of micropillars of pure Mg, Mg-4 at.%Al and Mg-9 at.%Al with different lateral cross-section (in the range 3x3 μm$^2$ to 7x7 μm$^2$) are plotted in Figs. 6a, b and c, respectively. The three regions analysed in the previous sections were found in all cases, regardless of the micropillar size and Al content. The critical stress for twin nucleation was higher in the 3x3 μm$^2$ micropillars but no differences were found between 5x5 μm$^2$ and 7x7 μm$^2$ micropillars.

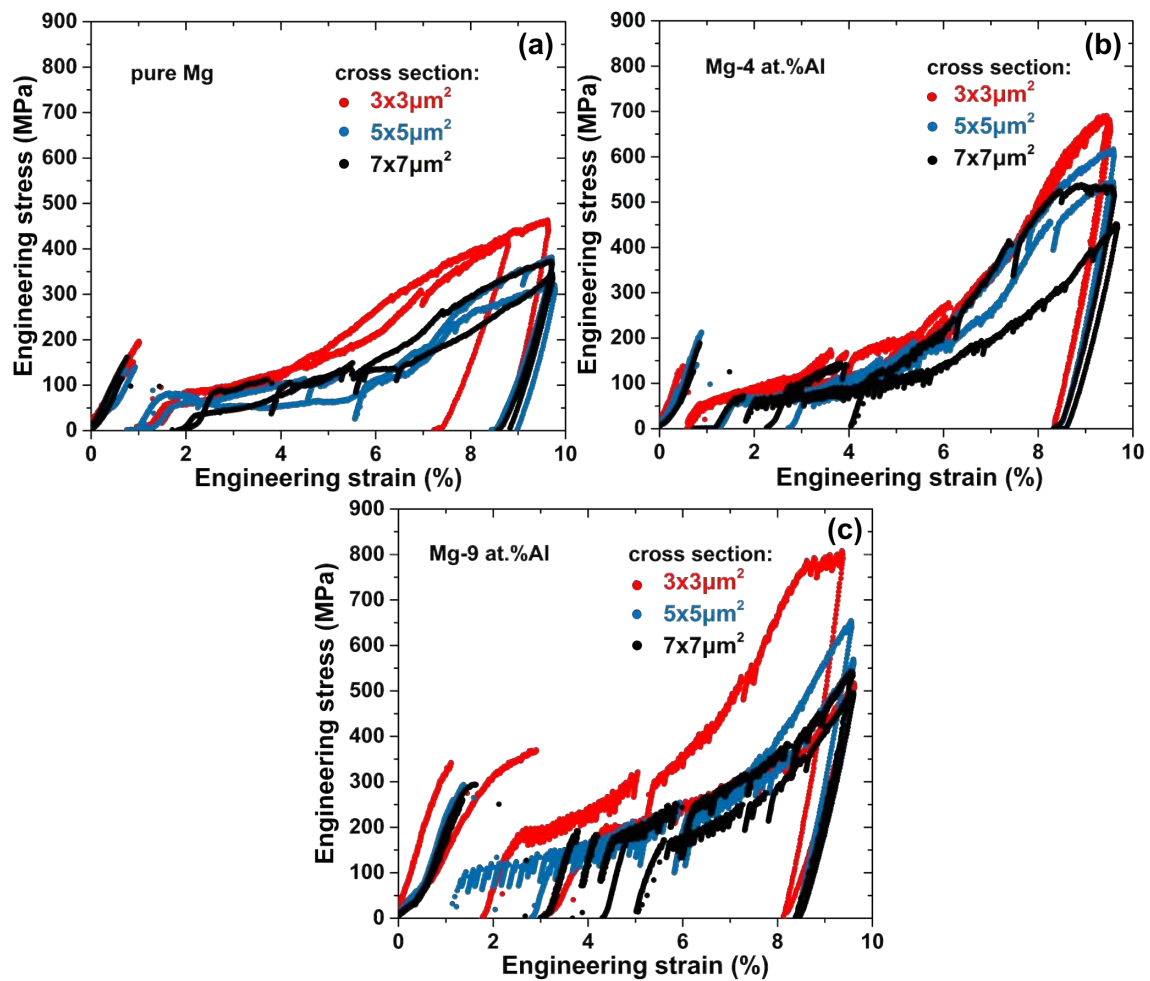

Fig. 6 Representative engineering stress-strain curves of micropillar compression tests along [01$\bar{1}$0] direction in micropillars with lateral cross section in the range 3x3 μm$^2$ to 7x7 μm$^2$. (a) Pure Mg, (b) Mg-4 at.%Al alloy and (c) Mg-9 at.%Al alloy.

The constant stress plateau after twin nucleation, that was associated to the propagation of one dominant twin variant, was again slightly higher in the 3x3 μm$^2$ micropillars of pure Mg, as compared with larger micropillars. However, these differences were not found in the Mg-Al micropillars. Finally, a size effect was also found in the strain hardening region for applied strains >5% in the case of 3x3 μm² micropillars of Mg and Mg-Al alloys. However, no significant differences were found in the strain hardening of the larger micropillars. It should also be noticed that the macroscopic shape of the micropillars after testing was independent of the micropillar size and Al content. Slip traces parallel to the basal plane of the twinned crystal were always present (Fig. 7) and everything indicates that the dominant deformation mechanisms were not modified by the micropillar dimensions.

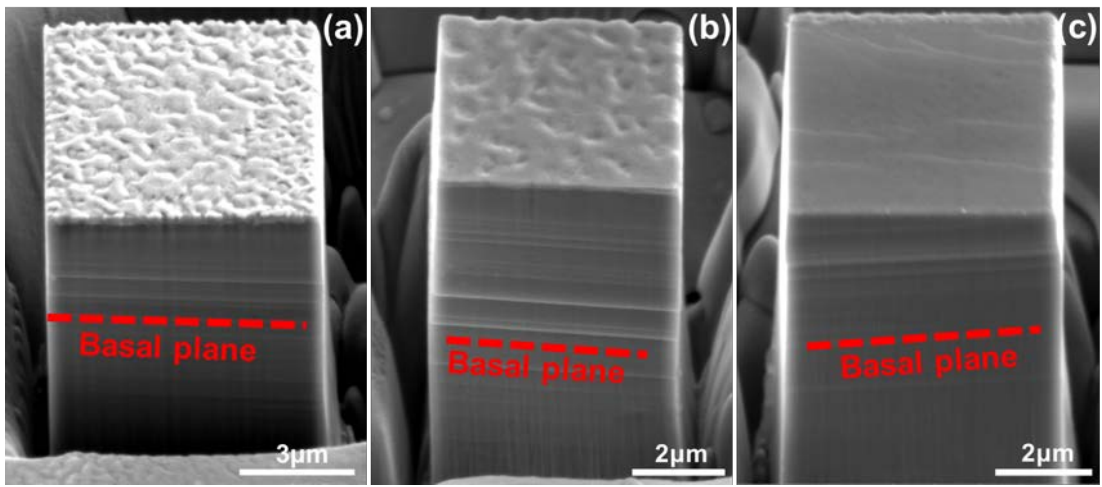

Fig. 7 SEM micrographs of the 5x5 μm² micropillars deformed up to 10%. (a) pure Mg, (b) Mg-4 at.%Al alloy and (c) Mg-9 at.%Al alloy. Slip traces along the basal plane of the twinned region are indicated by the red dash lines.

## 4. Discussion

### 4.1 Effect of the micropillar size on the CRSS for twin nucleation and growth

It is well-known that critical stress for dislocation slip [21,30–32,35,46] as well as twin nucleation and growth [17, 26, 28–31, 35, 47–50] increase as the micropillar size decrease. Thus, the analysis of this size effect is mandatory to assess the ability of the micropillar compression tests to provide accurate information about the influence of the solute atoms on the critical stress for twin nucleation and growth. The average values and standard deviation of the critical shear stresses for $\{10\bar{1}2\}$ twin nucleation, $\tau_n$, and growth, $\tau_g$, for pure Mg, Mg-4 at.%Al and Mg-9 at.%Al alloys are plotted as a function of micropillar size in Figs. 8(a) and 8(b), respectively. The twin nucleation stress ($\tau_n$) is defined as the resolved shear stress on the twin plane at the peak stress associated to the twin nucleation. The twin growth stress ($\tau_g$) stands for the shear stress resolved on the twin plane in the beginning of the plateau region, after the load drop due to the strain burst associated with the nucleation of the twin.

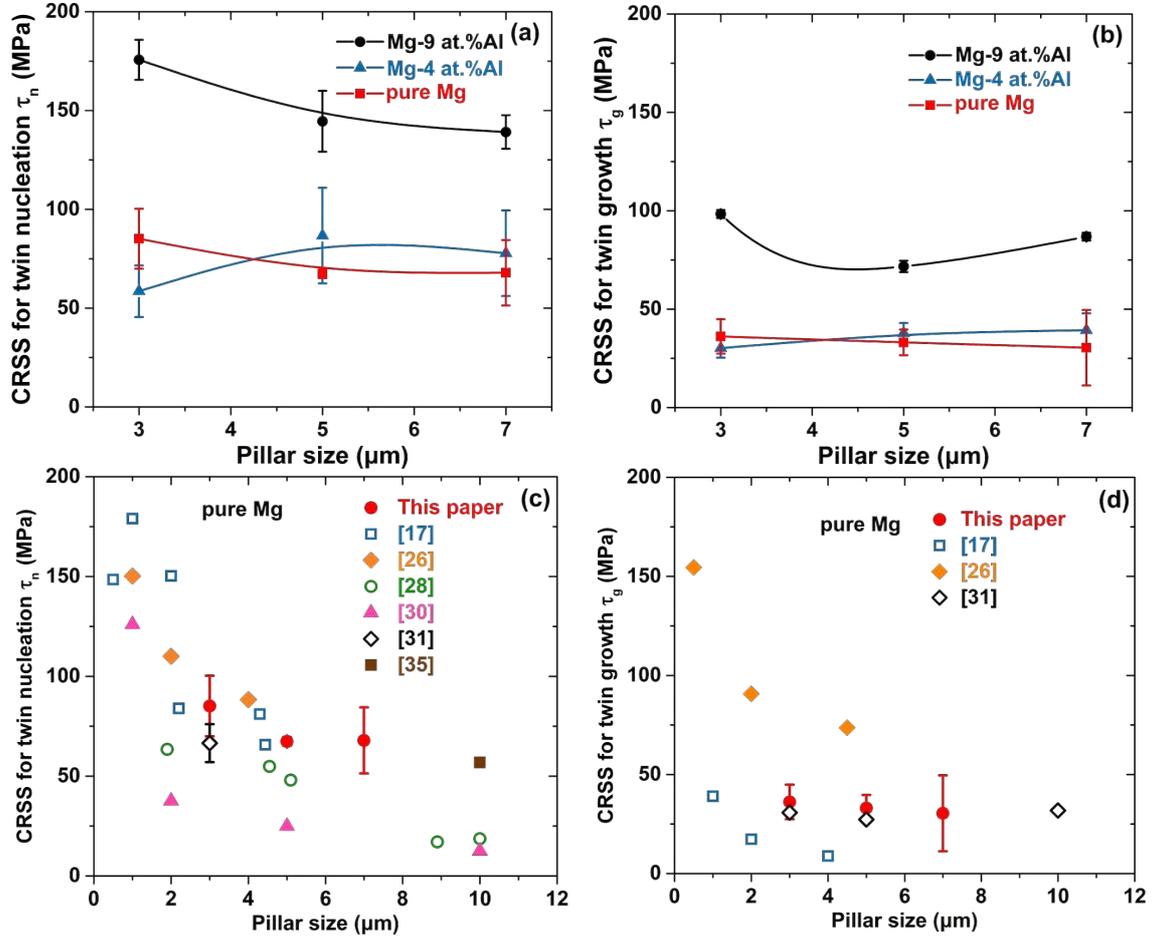

Fig. 8 (a) CRSS for twin nucleation ($\tau_n$) as a function of the micropillar lateral dimension and Al content. (b) CRSS for twin growth ($\tau_g$) as a function of the micropillar lateral dimension and Al content. (c-d) Effect of micropillar dimension on the CRSS for twin nucleation and growth for pure Mg. Results from this investigation and from the literature [17,26,28,30,31,35].

A size effect of the type 'smaller is stronger' was found in the CRSS for twin nucleation in the micropillars of 3x3 μm² of Mg and Mg-9 at.%Al. Nevertheless, the values of $\tau_n$ were similar for micropillars of 5x5 μm² and 7x7 μm², regardless of the Al content. In the case of the CRSS for twin growth, a size effect was only found in the micropillars of 3x3 μm² of Mg-9 at.%Al and no significant differences in $\tau_g$ were found for Mg and Mg-4 at.%Al as a function of the micropillar dimensions. In the case of pure Mg, the experimental values of $\tau_g$ and $\tau_n$ have been plotted in Figs. 8c and 8d as a function of the micropillar lateral dimension together with those available in the literature. They show large differences in the critical stress values among different investigations but the size effects for both twin nucleation and growth are normally limited to micropillars with lateral dimensions below 3 µm. The large differences in the CRSS can be attributed to the localized nature of the twin nucleation, which is triggered at the contact point between the flat punch and the micropillar upper surface. Small misalignments (which are very difficult to correct) can lead to large differences in the measured critical stress. Overall, the experimental data for pure Mg obtained in this investigation are consistent with most of the results in the literature.

**4.2 Effect of Al content on the CRSS for twin nucleation and growth**

As the effect of the micropillar dimensions on the critical shear stress for twin nucleation and growth was negligible for micropillars of 5x5 µm$^2$ and 7x7 µm$^2$ cross-section, the values measured with the latter were used to assess the effect of Al content on them. They are plotted in Fig. 9a and the CRSS for twin nucleation approximately 80 MPa higher than that for twin growth, regardless of the Al content. The hardening contribution due to the presence of Al atoms in the CRSS for twin nucleation and growth, expressed as $\Delta\tau_n = \tau_n(Al) - \tau_n(0)$ and $\Delta\tau_g = \tau_g(Al) - \tau_g(0)$, is plotted in Fig. 9b.

Twin thickening is a process mediated by the propagation of twinning dislocations and the solute strengthening was analysed by Ghazisaeidi et al. [51] using a modified version of Labusch model [52] that takes into account that a straight dislocation bows out in the glide plane to minimize the potential energy in the presence of a random solute distribution. The interaction energy between the solute atoms with the twin dislocation (and also with the twin boundary) was determined form first principles calculations, leading to a linear dependence of the CRSS with the Al content for a given value of the strain rate and temperature. The difference with the traditional Labusch model, that predicts that the yields scales with the solute atom concentration to the power of 2/3, is due to the small Burgers vector of the twinning dislocation and to the solute-twin boundary interaction, that provides the main contribution to the strengthening. The predictions obtained from eq. (8) in [51] for ambient temperature and a strain rate of 10$^{-3}$ s$^{-1}$ are plotted in Fig. 9b. They are in good agreement with the experimental results for Mg-4 at.%Al but underestimate the experiments for Mg-9 at.%Al. The origin of these differences may be found in that the energy interaction between the solute atoms and the twin dislocation and the twin boundary was calculated using a dilute approximation, that might not be accurate when the Al content is 9 at.%.

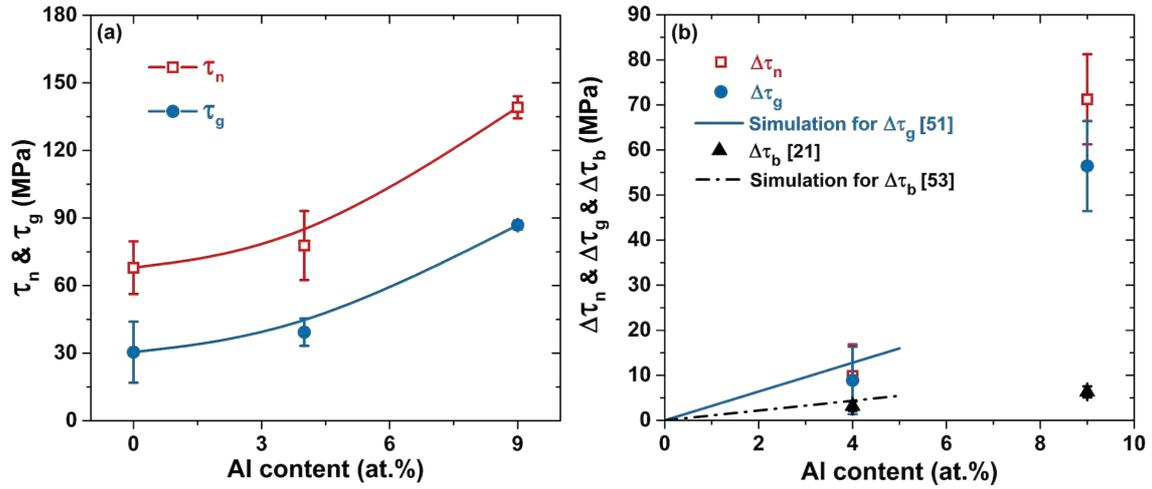

Fig. 9 (a) CRSS for twin nucleation, $\tau_n$, and growth, $\tau_g$, measured in micropillars of 7x7 μm² cross-section as a function of Al content. (b) Increase of the CRSS for twin nucleation, $\tau_n$, and growth, $\tau_g$, as a function of the Al content. Experimental data of the increase of CRSS for basal slip [21] and the first-principle predictions of the effect of Al content on the CRSS for twin growth [51] and basal slip [53] are also plotted for comparison.

The recent experimental data of the influence of Al content on the CRSS for basal slip $\Delta\tau_b$ [21], obtained using the same experimental approach, have been plotted in Fig. 9b as well, together with the predictions from [53] based on the modified Labusch model. The agreement between experiments and simulations is good up to 4 at. %Al and the extrapolation of the model to the non-dilute range also seems to be in agreement with the experimental results in the case of basal slip. Moreover, both experimental and simulations show that the strengthening provided by Al solute atoms is higher in the case of twin nucleation and growth than in the case of basal slip.

### 4.3 Strain hardening in the twinned micropillars

The stress-strain curves of the micropillars oriented for twinning showed a region with strong strain hardening when the applied strain was higher than 5%. Similar behaviour has been reported during micropillar compression [28,31] as well as uniaxial [54] and plane strain compression [55] Mg and Mg alloys single crystals oriented for twinning. The origin of this strain hardening has been attributed to the activation of <c+a> pyramidal slip to accommodate the deformation in the twinned crystal, which is not suitably oriented for basal slip. Moreover, other authors also pointed out the strengthening may also be caused by the presence of TTB which may block dislocation slip [31]. It should also be noted that the strain hardening rate increased rapidly with the Al content for all micropillar sizes.

After twinning, the compression direction was close to the *c*-axis and new micropillar compression tests were carried out in specimens of Mg-4 at.%Al and Mg-9 at.%Al alloys oriented in [0001] direction. There are 5 equivalent pyramidal slip systems in this orientation with high Schmid factors (between 0.46 and 0.43, see Table 1) which are similar to those in the [01$\bar{1}$6] orientation within the twinned region of the micropillars in Fig. 6 (between 0.37 and 0.49 in Table 1). Three tests were carried out for each alloy in micropillars of 7x7 µm$^2$ cross-section to ascertain the CRSS for <*c+a*> pyramidal slip as a function of the Al content. They were not performed in pure Mg because the CRSS for pyramidal slip is well established from previous studies [30,32]. Two representative resolved shear stress–strain curves obtained from these experiments are plotted in Fig. 10a. The initial response is elastic followed by a plastic region with linear strain hardening. This behaviour is equivalent to that found in pure Mg deformed in compression along the [0001] direction [30,32]. The analysis of the deformed micropillars in the SEM showed the presence of slip traces parallel to the basal plane (Fig. 11a), very similar to those found in the micropillars tested along [01$\bar{1}$0] after twinning (Fig. 7). The TKD map of a thin lamella extracted from that deformed pillar of Mg-4 at.%Al did not show any evidence of twinning (Fig. 11b), indicating that the plastic deformation was accommodated by dislocation slip.

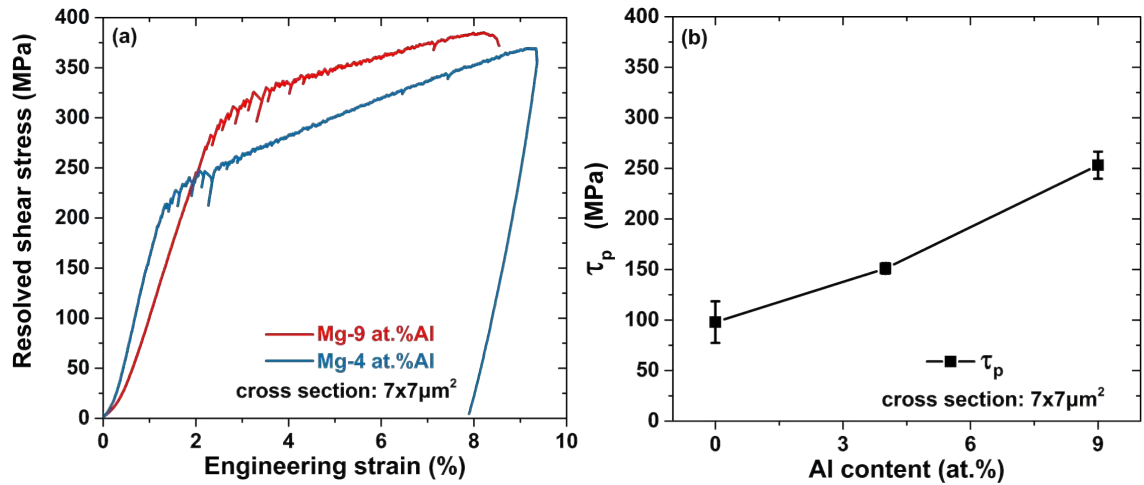

Fig. 10 (a) Representative resolved shear stress – e strain curves of micropillars of 5x5 and 7x7 µm$^2$ cross-section of Mg-4 at.%Al, and Mg-9 at.%Al. (b) CRSS to activate pyramidal dislocation slip in Mg, Mg-4 at.%Al and Mg-9 at.%Al. The results for pure Mg are obtained from [30,32].

The CRSS for the initiation of plastic slip could be easily computed from the onset of non-linear deformation in Fig. 10a and the average values (as well as the standard deviation) are plotted in Fig. 10b as a function of the Al content. The experimental results in the case of pure Mg included in this figure are taken from [30,32]. The dislocation structure in the deformed micropillars was analysed by TEM. As found within the twin region of the micropillars deformed along [01$\bar{1}$0], <*a*> prismatic dislocations as well as <*c+a*> pyramidal dislocations were found under the diffraction conditions *g*=[0001] (Fig. 11c) and *g*=[01$\bar{1}$0] (Fig. 11d), respectively. Thus, plastic deformation along [0001] was mainly accommodated by the pyramidal slip, and the influence of the Al content on the CRSS for pyramidal slip can be found in Fig. 10b.

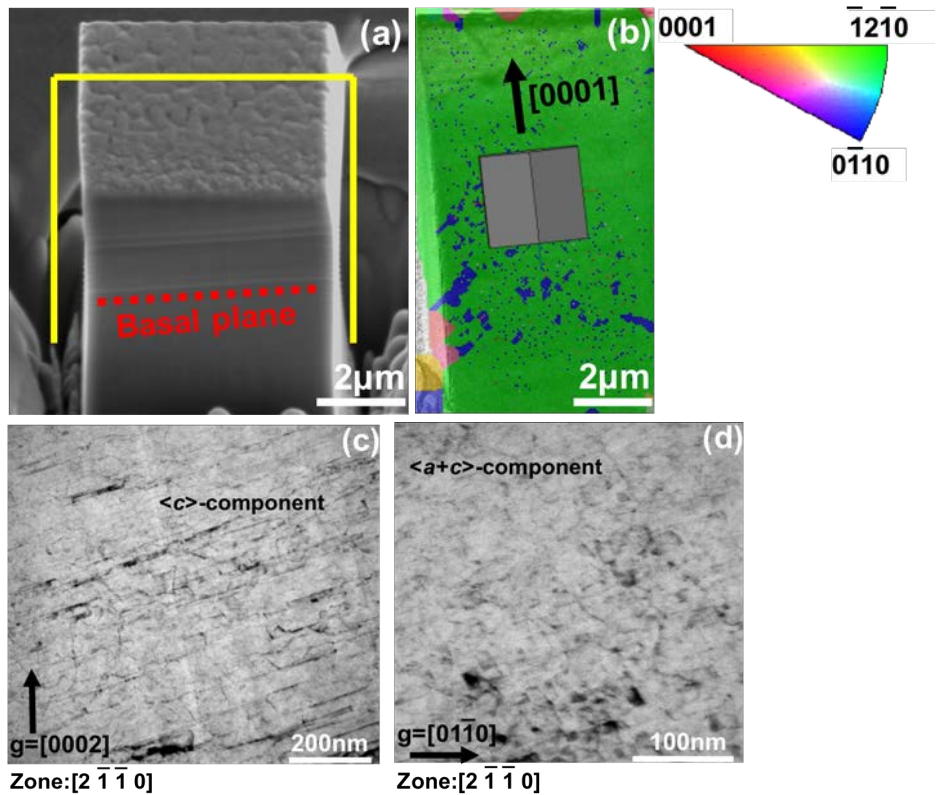

Fig. 11 (a) SEM micrograph of the Mg-4 at.%Al micropillar deformed along the [0001] direction. (b) *t*-EBSD maps of a thin lamella extracted from the micropillar in (a). (c) TEM micrographs of that lamella under the condition *g*=[0002] to reveal the <*c*> dislocations. (d) Same as (c) observed under *g*=[01$\bar{1}$0] to reveal <*a*> dislocations.

The presence of solute atoms increased dramatically the CRSS for pyramidal slip ($\tau_p \approx$ 98 MPa in pure Mg to $\tau_p \approx$ 250 MPa in Mg-9 at.%Al) and these differences may explain the effect of the Al content of the strain hardening rate of the micropillars deformed along [01$\bar{1}$0] after twinning. In the case of pure Mg, plastic deformation by pyramidal slip is activated when the CRSS on the pyramidal system reaches around 100 MPa, and this stress level is attained at low strains when the applied engineering stress in the micropillar reaches ≈ 200 MPa (≈6% in Fig. 6a). Thus, most of the strain hardening of the twinned Mg micropillar up to 10% strain is associated to the interaction between pyramidal dislocations and also the interaction of pyramidal dislocations with TTB [30,32]. Activity of pyramidal dislocations was very large under these conditions, as shown in Figs. 3c and 3d. Nevertheless, the CRSS to activate pyramidal slip in the twinned Mg-9 at.%Al micropillars was much higher (250 MPa) and compressive stress higher than 500 MPa were necessary to promote pyramidal slip. These stresses were only achieved at very high strains (> 8%) in the micropillars with 5x5 and 7x7 µm$^2$ cross-section. Thus, in the absence of pyramidal slip and with very limited basal slip activity, the deformation has to be partially accommodated by elastic strains, leading to very large strain hardening rates in the Mg-Al alloys, as compared with pure Mg. This mechanism was confirmed by limited presence of *<c+a>* pyramidal dislocations in the Mg-Al micropillars after 10% strain (Figs. 5b and c).

## 5. Conclusions

The effect of Al content on the critical resolved shear stress for twin nucleation and growth was analysed by the combination of diffusion couples with micropillar compression tests. Micropillars of square cross cross-sections in the range 3x3 µm$^2$ to 7x7 µm$^2$ were deformed along [01$\bar{1}$0]. The deformation mechanisms were independent of the micropillar dimensions and Al content. The stress for twin nucleation was identified by a sudden strain burst due to the nucleation of one or two twin variants in the upper region of the micropillar. Afterwards, twin propagation occurred at a constant stress until the whole micropillar was twinned and deformation was partially accommodated by deformation of the twinned micropillar by pyramidal and basal slip. These mechanisms were accompanied by a strong strain hardening.

Size effects for micropillars of 5x5 µm$^2$ to 7x7 µm$^2$ were negligible and the critical resolved shear stresses for twin nucleation and growth could be obtained from the applied stress and the Schmid factors. The critical resolved shear for twin nucleation was higher than that for twin growth both increased by the same amount with the Al content. Nevertheless, the increase was small (≈ 10 MPa) for 4 at.%Al and large (up to 60-70 MPa) for 9 at.%Al. The solute strengthening in the dilute regime (< 5 at.%Al) was in agreement with the predictions of Labusch-type models based of first principles calculations. However, this approach underestimated the strengthening for large solute contents. In addition, it was shown Al atoms are more effective in increasing the critical resolved shear stresses for twin nucleation and growth than for basal slip.

Compression tests in micropillars oriented along [0001] showed that the strain hardening in the twinned micropillars was linked to pyramidal slip. It was found that the critical shear stress for pyramidal slip increased rapidly with the Al content from 98 MPa in pure Mg to 250 MPa in Mg-9 at.%Al. Thus, the information about the critical resolved shear stress for twinning and pyramidal slip in Mg-Al obtained in this investigation, together with the corresponding results for basal slip in [21] show that the plastic anisotropy of Mg alloys increases with the Al content.

**Acknowledgements**

This investigation was supported by the European Research Council (ERC) under the European Union's Horizon 2020 research and innovation programme (Advanced Grant VIRMETAL, grant Agreement no. 669141). Ms. J-Y. Wang acknowledges the financial support from the China Scholarship Council, grant no. 201506890002.